\documentclass{PoS}
\usepackage{amssymb}
\usepackage{amsmath}
\usepackage{textcomp}
\usepackage[utf8]{inputenc}
\usepackage{upgreek}
\usepackage{lineno}

\def\bi{\begin{itemize}}
\def\ei{\end{itemize}}
\newcommand{\sub}[1]{\ensuremath{_{\text{{#1}}}}}

\title{GAPS -- Dark matter search with low-energy cosmic-ray
   antideuterons and antiprotons}

\ShortTitle{GAPS -- Dark matter search with low-energy cosmic-ray
   antideuterons and antiprotons}

\author{\speaker{P. von Doetinchem}\\
        University of Hawai'i at M$\bar{\text{a}}$noa\\
        E-mail: \email{philipvd@hawaii.edu}}
\author{T. Aramaki, C.J. Hailey\\
Columbia University}       
        
\author{S. Boggs\\
University of California at Berkeley}

\author{H. Fuke\\
Institute of Space and Astronautical Science, Japan Aerospace Exploration Agency}

\author{S.I. Mognet, R.A. Ong, J. Zweerink\\
University of California at Los Angeles}

\author{K. Perez\\
Haverford College}

\abstract{The GAPS experiment is foreseen to carry out a dark matter search by measuring low-energy cosmic-ray antideuterons and antiprotons with a novel detection approach. It will provide a new avenue to access a wide range of different dark matter models and masses from about 10\,GeV to 1\,TeV. The theoretically predicted antideuteron flux resulting from secondary interactions of primary cosmic rays is very low. Well-motivated theories beyond the Standard Model contain viable dark matter candidates, which could lead to a significant enhancement of the antideuteron flux due to annihilation or decay of dark matter particles. This flux contribution is believed to be especially large at low energies, which leads to a high discovery potential for GAPS. The GAPS low-energy antiproton search will provide some of the most stringent constraints on \texttildelow30\,GeV dark matter, will provide the best limits on primordial black hole evaporation on galactic length scales, and explore new discovery space in cosmic-ray physics.  

GAPS is designed to achieve its goals via long duration balloon flights at high altitude in Antarctica. The detector itself will consist of 10 planes of Si(Li) solid state detectors and a surrounding time-of-flight system. Antideuterons and antiprotons will be slowed down in the Si(Li) material, replace a shell electron and form an excited exotic atom. The atom will be deexcited by characteristic X-ray transitions and will end its life by the formation of an annihilation pion/proton star. This unique event structure will deliver a nearly background free detection possibility. 
}

\FullConference{The 34th International Cosmic Ray Conference,\\
		30 July- 6 August, 2015\\
		The Hague, The Netherlands}

\begin{document}
\clubpenalty = 10000 \widowpenalty = 10000 \displaywidowpenalty = 10000

\section{Identification of dark matter}

The existence of dark matter is established on very different length scales from galaxies to galaxy clusters to the cosmic microwave background~\cite{darkmatter}. The importance of this problem becomes obvious by recalling that dark matter is more than five times more abundant than regular matter. Little is known about the nature of dark matter particles except that they are gravitationally interacting, but other than that only very weakly interacting -- if at all -- with regular matter. Many theories trying to explain dark matter provide a stable, relatively heavy particle -- the weakly interacting massive particle (WIMP). All WIMPs invoke physics beyond the standard model of particle physics. It is generally acknowledged that the identification of dark matter will require an interplay of different types of instruments, including collider experiments (not further discussed). There are about ten operating or planned experiments to detect WIMPs through their recoil off of target nuclei~\cite{baudis}, which effectively probes a scattering cross section (direct searches). So far, the current flagship direct searches are most sensitive to intermediate-mass (\texttildelow100\,GeV) dark matter and it might remain challenging to conclusively resolve the light dark matter discrepancies~\cite{fox}. 

If dark matter was in thermal equilibrium with the rest of the matter in the early universe it is a natural assumption that dark matter particles are able to interact with each other and produce known Standard Model particles. These particles would contribute to the known cosmic-ray fluxes, and thus it could be possible to observe an imprint of dark matter in the diffuse cosmic-ray spectra in the form of an excess (indirect searches). This process is complementary to the scattering approach of the direct searches, and thus probes different parameter regions of dark matter theories. Cosmic-ray antiparticles without primary astrophysical sources are ideal candidates for an indirect dark matter search, but recent results show that accomplishing this task with positrons and higher energy antiprotons appears to be challenging due to high levels of secondary/tertiary astrophysical background. However, the latest results of major cosmic-ray instruments (e.g., AMS-02~\cite{ams4}) for the electron and positron data show evidence of a structure that might be interpreted as dark matter. Also antiprotons have been used to constrain a variety of dark matter models, e.g., \cite{2015arXiv150404276G}, though their statistical accuracy at low energies is limited. On the $\upgamma$-ray front, Fermi has used stacked measurements of dwarf spheroidal galaxies and measurements of the extragalactic $\upgamma$-ray background to constrain the dark matter annihilation cross section, e.g.~\cite{Ackermann:2015tah}. In 2014, a $\upgamma$-ray excess observed by Fermi in the Galactic Center region was interpreted as originating from \texttildelow30\,GeV dark matter~\cite{Daylan:2014rsa}. The excess shows a high significance and interpretation is ongoing. 

\subsection{Exploring New Phase Space: Low-energy Antideuterons}

\begin{figure}
\centering
\includegraphics[height=0.4\linewidth]{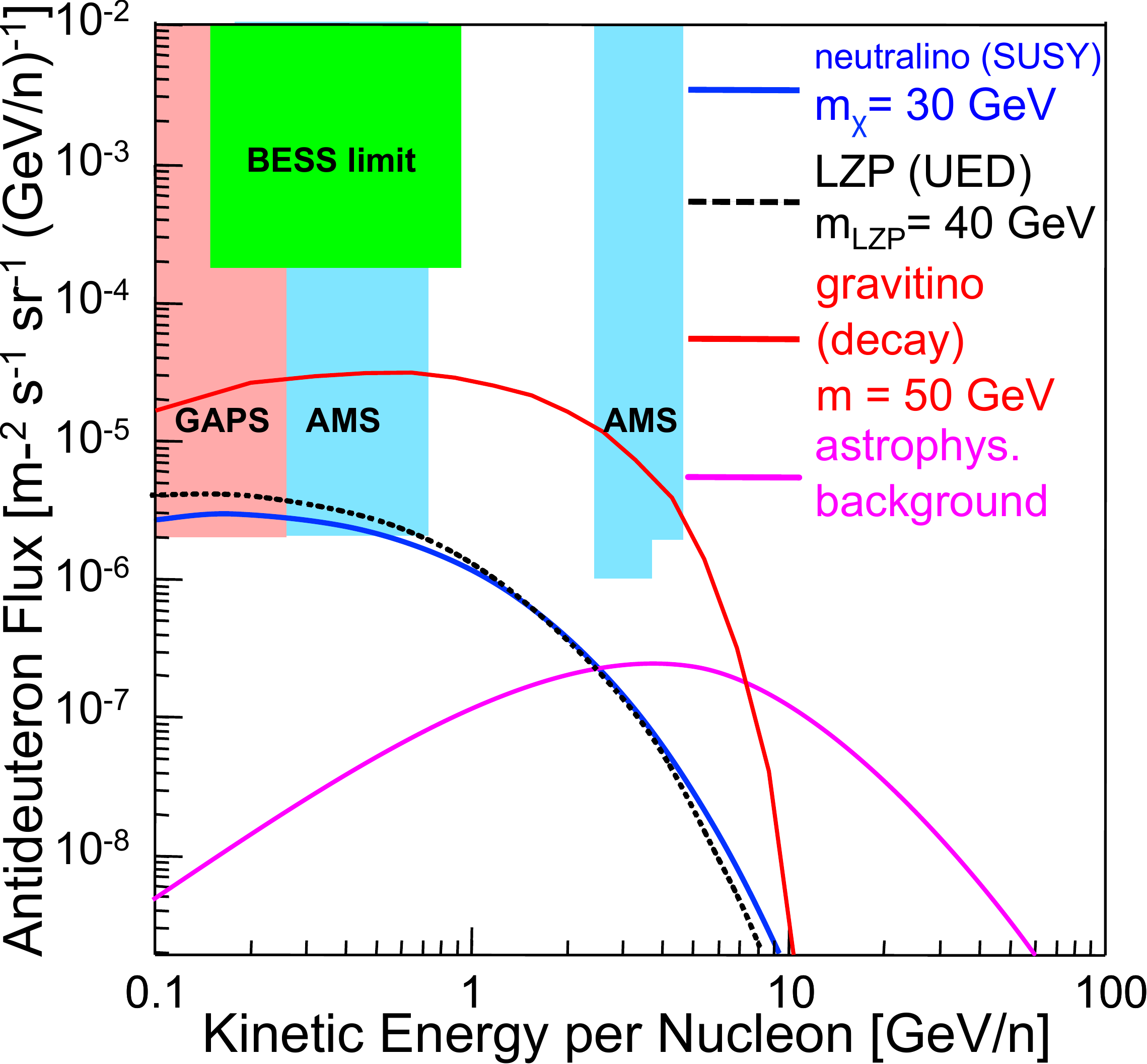}
\hspace{.1\linewidth}
\includegraphics[height=0.39\linewidth]{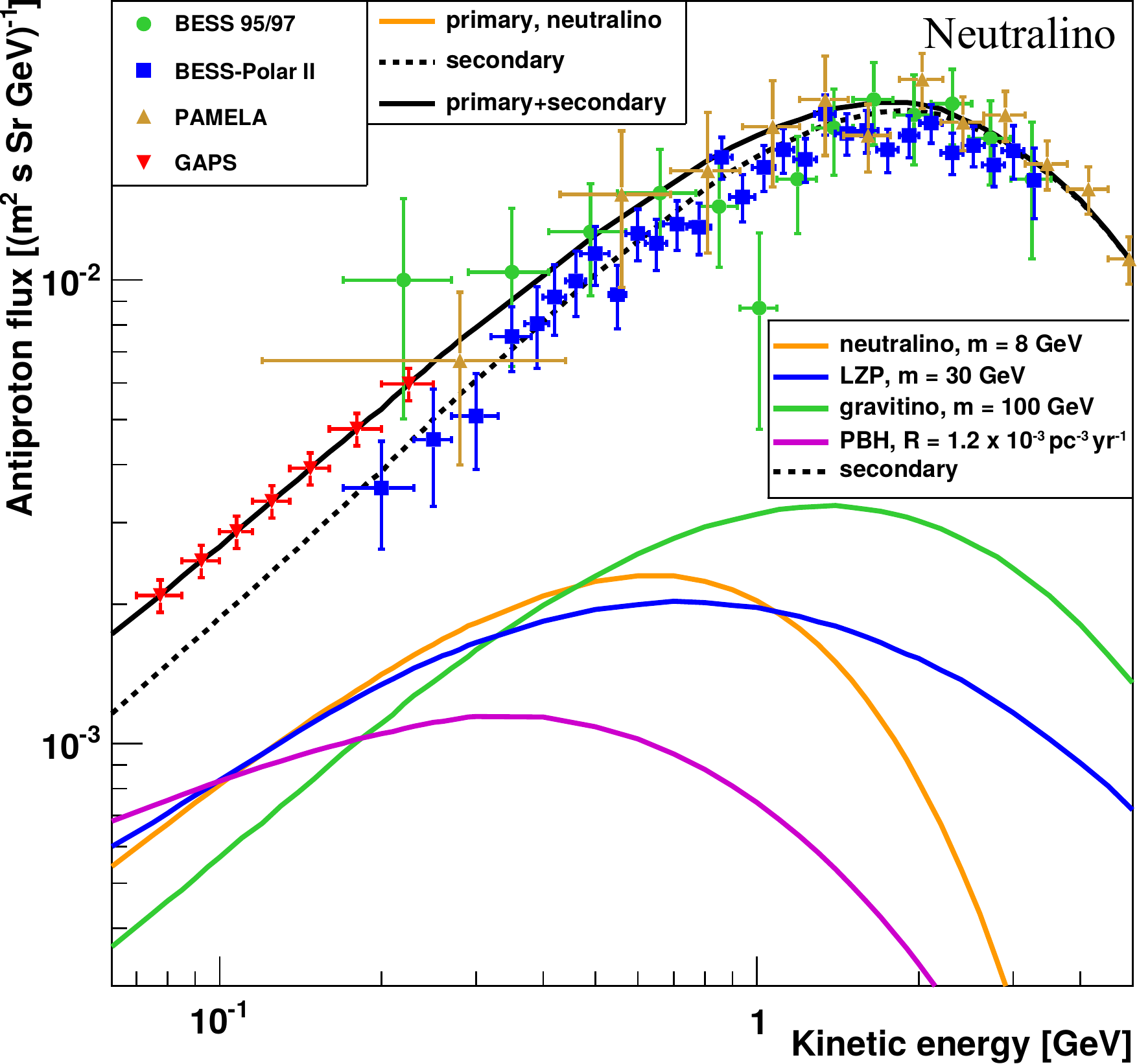}
\caption{\textit{Left)} Antideuteron limits from BESS \cite{Fuke:2005it}, predicted antideuteron fluxes from different models \cite{Baer:2005tw,Donato:2008yx,Dal:2014nda,Ibarra:2013qt}, sensitivities for the running AMS-02 for 5\,years \cite{ams02dbaricrc2007} and the planned GAPS experiments after three 35-day flights \cite{gaps,Aramaki2015}. \textit{Right)} Predicted primary antiproton fluxes at TOA from neutralinos, LZPs, gravitinos, or PBHs, along with neutralino signal as seen by 1~GAPS LDB flight~\cite{Aramaki:2014oda}. \label{f-dmdbar}}
\end{figure}

Antideuterons may be generated in dark matter annihilations or decays, offering a potential breakthrough in unexplored phase space for dark matter. A low-energy search for antideuterons has ultra-low astrophysical background~\cite{2015arXiv150507785A}. Secondary/tertiary (background) antideuterons, like antiprotons, are produced when cosmic-ray protons or antiprotons interact with the interstellar medium, but the production threshold for this reaction is higher for antideuterons than for antiprotons. Collision kinematics also disfavor the formation of these low-energy antideuterons. Moreover, the steep energy spectrum of cosmic rays means there are fewer particles with sufficient energy to produce secondary antideuterons, and those that are produced have relatively large kinetic energy. The left panel of Fig.~\ref{f-dmdbar} shows the antideuteron flux expected from three benchmark dark matter scenarios (further discussed in Sec.~\ref{s-pot}) using the MED propagation model~\cite{2004PhRvD..69f3501D} compared to the expected secondary/tertiary background~\cite{Ibarra:2013qt}. The figure reveals why low-energy antideuterons are such an important approach: the flux from dark matter interactions exceeds the background flux by more than two orders of magnitude in the energy range $<$0.25\,GeV/$n$ without relying on any boosting mechanisms, e.g., due to dark matter clumpiness, Sommerfeld enhancement, or large galactic halo size. Introducing such effects, the signal-to-background ratio would further increase. This is in contrast to dark matter signal predictions for positrons and higher energy antiprotons, which are typically only a small additional contribution on top of the astrophysical background. 

The uncertainties of antideuteron propagation and coalescence are both on the order of $\mathcal{O}(10)$~\cite{2015arXiv150507785A}. It is important to stress that recent observations and calculations disfavor dark matter halo models with a small size (MIN model). Therefore, the fluxes shown in the left side of Fig.~\ref{f-dmdbar} are a lower limit in the MED model and would be higher in the MAX model. In addition, the antideuteron fluxes presented here assume a very conservative boost factor due to dark matter clumps of $f=1$. However, a factor of $f=2$--3, consistent with current theoretical expectations, would shift dark matter fluxes up by a factor $f$, increasing the GAPS reach into discovery space~\cite{Baer:2005tw,Lavalle:1900wn}.

\section{Discovery potential of the General Antiparticle Spectrometer (GAPS)\label{s-pot}}

The General Antiparticle Spectrometer (GAPS) is a large-acceptance cosmic-ray experiment, designed to measure low-energy antideuterons and antiprotons during Antarctic long-duration balloon (LDB) flights~\cite{gaps}. The GAPS antideuteron and antiproton searches will provide a new avenue to access a wide range of dark matter models and dark matter masses from 10\,GeV to 1\,TeV~\cite{2015arXiv150507785A}. It is important to stress that GAPS uses an entirely different particle identification technique than the magnetic spectrometers AMS-02 and BESS, allowing for detailed studies of systematic uncertainties. Sec.~\ref{s-1ldb} and \ref{s-3ldb} highlight some of the accessible dark matter models for one and three LDB flights. The instrumental design and identification technique is further discussed in Sec.~\ref{s-det}. 

\subsection{Potential for one LDB flight\label{s-1ldb}}

One GAPS LDB flight will permit a deep search for dark matter using antideuterons, improving the existing antideuteron limits given by BESS~\cite{Fuke:2005it} by \texttildelow1.5 orders of magnitude. Already one GAPS flight will effectively test models with decaying dark matter. These models provide a very long-lived unstable dark matter candidate, which decays into Standard Model cosmic rays including antideuterons and antiprotons~\cite{Dal:2014nda}. A particularly well-motivated and well-studied case of decaying dark matter is the gravitino in supergravity models with $R$-parity violation. Proton stability is not an issue as long as only specific subsets of $R$-parity violating operators are present in the theory. If the gravitino decays via certain trilinear $R$-parity violating operators the expected antideuteron flux is enhanced compared to the antiproton flux and allows for detection with one GAPS LDB flight. A recently updated flux in agreement with antiproton bounds for a 50\,GeV gravitino is shown in the left side of Fig.~\ref{f-dmdbar}~\cite{Dal:2014nda}. As mentioned above, all models shown here are using a conservative boost factor of $f=1$. Relaxing this to a still-reasonable value of $f=3$, one LDB flight would yield equivalent sensitivity to three LDB flights with $f = 1$. 

Also with one LDB flight, GAPS will open up a new field of precision antiproton measurements by detecting \texttildelow2 orders of magnitude more low-energy antiprotons than BESS and PAMELA combined~\cite{2012PhRvL.108e1102A,pamelapbar}. At kinetic energies below 0.25\,GeV, the statistical accuracy of the BESS and PAMELA data that have been used to constrain dark matter models is very low. Since the antiproton spectrum from dark matter annihilations shifts towards lower kinetic energies with decreasing dark matter mass, the precision measurement of this energy range provided by GAPS offers new phase space for probing light dark matter models~\cite{Aramaki:2014oda}. The right side of Fig.~\ref{f-dmdbar} illustrates that the GAPS ultra-low energy antiproton measurement will be sensitive to light neutralinos, gravitinos, and a right-handed Kaluza-Klein neutrino of warped 5-dimensional grand unified theories (GUT) with a conserved $Z_3$ parity (LZP) for a reasonable range of propagation parameters. The GAPS antiproton search will also provide detailed studies of solar modulation, geomagnetic deflection, atmospheric interactions, all of which are essential for understanding cosmic-ray antiproton and antideuteron results. 

In addition to dark matter identification, low-energy antideuterons and antiprotons can constrain early universe physics. Density fluctuations, phase transitions, or the collapse of cosmic strings in the early universe may have formed primordial black holes (PBHs), which could have a sufficient evaporation rate to be observed now~\cite{Barrau:2002ru}. Antiprotons are produced as a result of jet fragmentation in the evaporating PBH. Calculations predict that the primary antiproton flux could be as high as the secondary flux at low energy, providing the best upper limit on the explosion rate of PBHs on a galactic scale. A strong constraint on PBH physics will be provided by the precision low-energy antiproton measurement from one flight. 

\subsection{Potential for three LDB flights\label{s-3ldb}}

Three GAPS LDB flights will probe even deeper into  beyond the Standard Model phase space. The left side of Fig.~\ref{f-dmdbar} shows the antideuteron spectrum from a viable supersymmetric neutralino with a mass of 30\,GeV annihilating into $b\bar b$. It clearly demonstrates that it is within the detectable range of GAPS. This model is especially interesting as it was recently suggested to explain the excess in the diffuse $\upgamma$-ray spectrum at the Galactic Center measured by Fermi~\cite{Daylan:2014rsa}. The same figure also illustrates the case of the aforementioned LZP case for the antideuteron channel~\cite{Baer:2005tw}. Furthermore, also gluonic annihilation channels of dark matter can produce a flux within the sensitivity of GAPS for masses below $\mathcal{O}(100\,\text{GeV})$~\cite{Cui:2010ud} and light candidates in the next-to-minimal supersymmetric model (NMSSM) still evade direct and collider bounds, but would be probed by antideuterons~\cite{Cerdeno:2014cda}. Antideuteron signals from heavy dark matter (5--20\,TeV) annihilation into $b\bar b$ might also be detectable by GAPS for the case of the MAX propagation model~\cite{Brauninger:2009pe}. Another option for heavy supersymmetric dark matter is pure-Wino dark matter~\cite{Hryczuk:2014hpa}. The thermal relic dark matter density is reached for Wino dark matter at masses in the TeV range, which might be detectable if Sommerfeld enhancement would be realized. In summary, the models briefly discussed above illustrate the broad range of dark matter candidates that GAPS is sensitive to.

\subsection{Complementarity with other search methods}
The use of multiple, complementary experiments has been successfully employed by the direct detection community. Separate experimental designs yield different backgrounds and approaches to suppressing these backgrounds, allowing for independent confirmation of any observed signal. As a rare event search, the hunt for cosmic-ray antideuterons requires a similar approach. Currently, this search relies exclusively on AMS-02~\cite{ams02dbaricrc2007}. GAPS and AMS-02 have complementary energy ranges, but some overlap at low energy, allowing the study of a large energy range for confirming signals and the best chance for controlling systematic effects. AMS-02, along with BESS and PAMELA, which provide the current best low-energy antiproton measurements, rely on magnetic spectrometers, and thus face different backgrounds than the GAPS exotic atom approach.

\section{The GAPS detector\label{s-det}}

\begin{figure}
\centering
\includegraphics[width=0.4\linewidth]{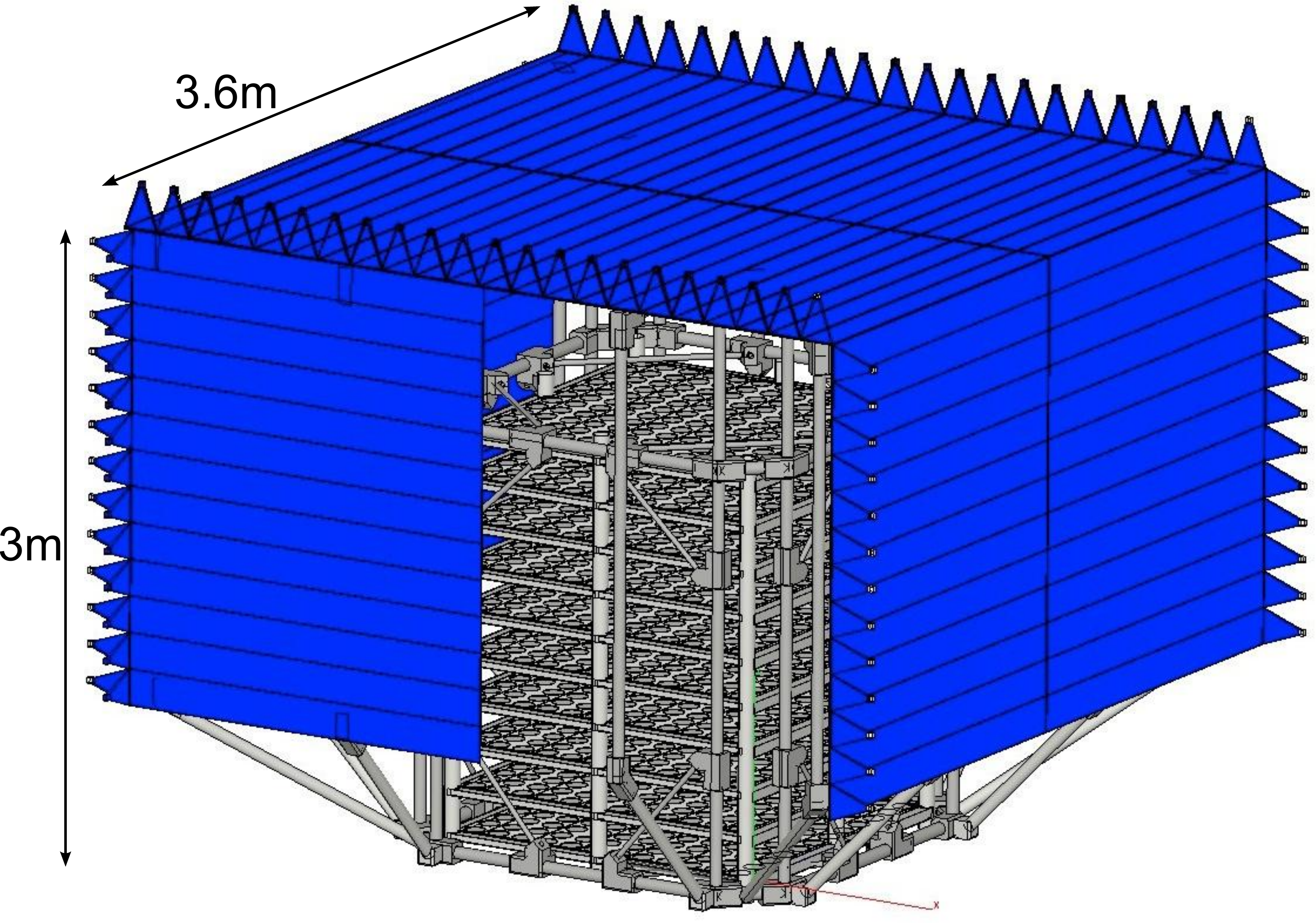}
\hspace{.1\linewidth}
\includegraphics[width=0.4\linewidth]{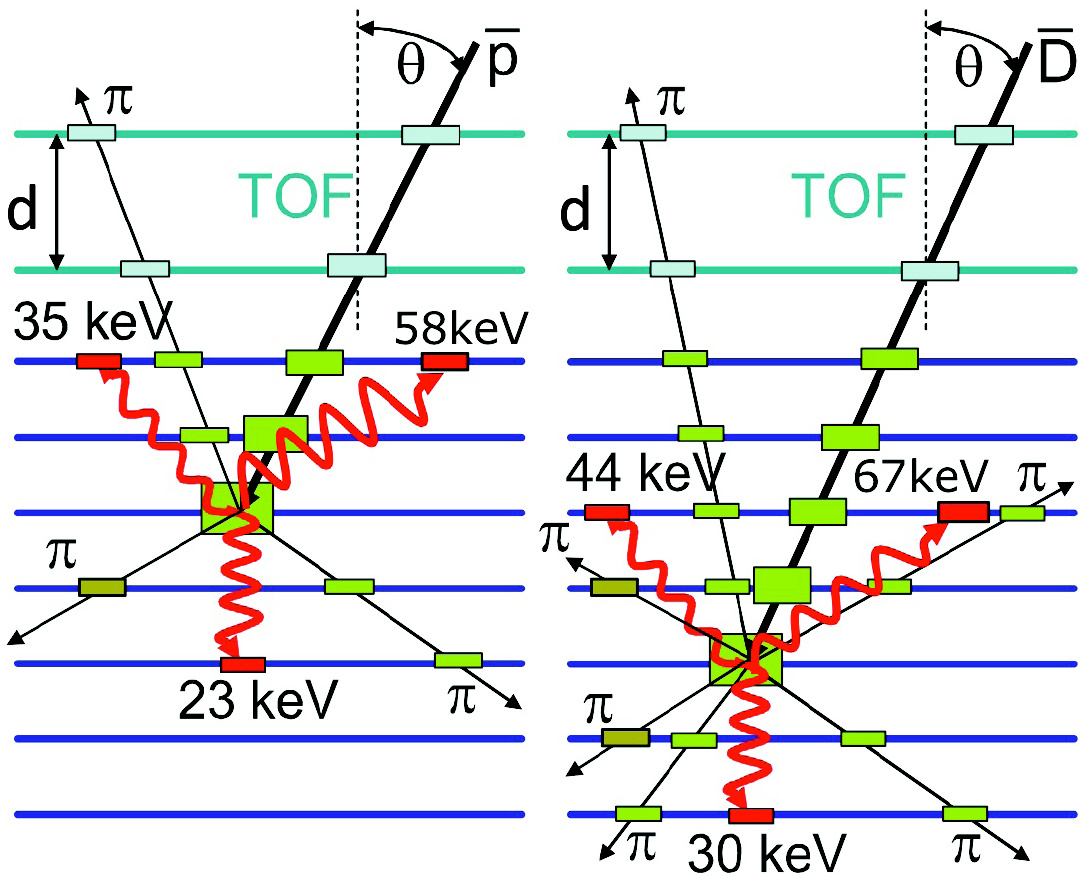}
\caption{\textit{Left)} TOF system and support structure with TOF paddles removed to expose Si(Li). \textit{Right)} For the same measured TOF and angle (i.e., particle velocity), an antideuteron (right) will penetrate deeper, typically emit twice as many annihilation pions and protons and emit X-rays of different well-defined energies than an antiproton (left). \label{f-layout}}
\end{figure}

The GAPS experiment (Fig.~\ref{f-layout}, left) is designed to reliably identify low-energy antideuterons and antiprotons during several Long Duration Balloon (LDB) campaigns from Antarctica. It utilizes a tracker made of 10\,planes of lithium-drifted silicon (Si(Li)) strip detectors in alternating orthogonal planes (Sec.~\ref{s-sili}), which is surrounded by a segmented plastic scintillator time-of-flight (TOF) system (Sec.~\ref{s-tof}). As discussed in the following, GAPS yields a large acceptance compared to experiments using magnets. In June 2012, a successful prototype balloon flight (pGAPS) from the balloon base in Taiki, Japan was carried out~\cite{2014NIMPA.735...24M,Doetinchem2013pGAPS}.

\subsection{Antiparticle identification}

An antideuteron or antiproton that has been slowed by the atmosphere passes through the TOF system (which measures particle velocity) and is slowed down and stopped by $\text dE/\text dx$ losses in the Si(Li) tracker. After stopping, the antiparticle forms an exotic atom in an excited state, with near unity probability. For the higher energy levels, the exotic atom de-excites through auto-ionizing transitions, and for the lower levels by radiative transitions. The radiative transition energies depend on the antiparticle-target atom reduced mass and the respective atomic numbers, and are uniquely predicted by quantum theory. The lower-energy transitions are in the \texttildelow10--100\,keV range. After X-ray emission, the antiparticle annihilates in the nucleus producing a nuclear star of charged pions (\texttildelow5.1 (3.1) for antideuterons (antiprotons)) and protons (\texttildelow1.8 (0.9) for antideuterons (antiprotons)). Each X-ray line energy uniquely identifies the antiparticle mass and charge. This is very important to reliably separate antideuterons from antiprotons (Fig.~\ref{f-layout}, right). A GAPS prototype was tested at the KEK accelerator in Japan in 2004/5~\cite{Hailey:2005yx}. The measurements confirmed the very high X-ray yields theoretically expected (but never before measured) from solid targets. In addition, measurements of X-rays and stars from antibaryons confirmed event topologies.  Recently a more detailed model~\cite{Aramaki2013}, applicable to determining X-ray yields for any antiparticle and target, was developed and validated with the results from KEK and other exotic atom studies. This model forms the basis of GAPS sensitivity estimates and predicts X-ray yields in silicon of \texttildelow75\%. The simultaneous occurrence in a narrow time window of X-rays of measured energy and a nuclear star with measured multiplicity provides an enormously constraining signature to identify antiparticle mass and charge and to suppress non-antiparticle background. More discrimination power to distinguish antideuterons from antiprotons comes from the fact that the particles GAPS is focusing on are significantly slower than minimum-ionizing particles. As the $\text d E/\text dx$ energy loss depends on kinetic energy, antideuterons of the same velocity as antiprotons penetrate about twice as deep into the Si(Li) tracker. Therefore, penetration depth together with velocity and change of $\text d E/\text dx$ from layer-to-layer provides additional rejection power. To identify slow antiprotons, the time-of-flight measurement and the simultaneous detection of two pions that are tracked back to a nuclear annihilation vertex are sufficient, while the detection of atomic X-rays can further constrain background events. Details of the antideuteron and antiproton identification and sensitivity are discussed in \cite{Aramaki2015,Aramaki:2014oda}.

\subsection{Si(Li) tracker\label{s-sili}}

Each plane of the Si(Li) tracker consists of about 140 circular Si(Li) detectors with 4"-diameter. Si(Li) detectors, with their superior energy resolution, deep depletion region and excellent timing resolution, are ideal for GAPS. The low atomic weight of silicon reduces internal background. At 0.25\,cm thickness, the Si(Li) detectors have high escape fractions down to \texttildelow20\,keV, well below the softest antiproton and antideuteron X-rays. In this way, Si(Li) serves as both target and detector, maximizing both the antideuteron stopping and X-ray detection efficiencies. Each 4" detector is segmented into four strips of equal area. The Si(Li) pixel size is optimized to save power and mass while providing sufficient spatial resolution for accurate track/depth reconstruction, so that the probability of more than one X-ray or particle track in the same strip is negligible. 

Detecting X-rays and particle tracks requires that the readout performs both \textit{1)} high-resolution spectroscopy on low-energy X-rays (\texttildelow10-100\,keV) and \textit{2)} particle tracking with coarse spectroscopy for the nuclear products ($\text dE/\text d x\approx1$--50\,MeV). The GAPS tracker will thus require the same dual-mode detection of events in the Si(Li) as used on pGAPS \cite{2014NIMPA.735...24M,Doetinchem2013pGAPS}. The detectors and electronics are designed to yield \texttildelow4\,keV (FWHM) resolution for the X-ray channel at temperatures of $-35\,^{\circ}${\rm C} to be able to resolve the different X-rays from antideuterons and antiprotons .

An Oscillating Heat Pipe (OHP) cooling system will be used to cool the Si(Li) detectors to their operating temperatures. An OHP was successfully flight-tested on pGAPS. A coupled thermal and fluid flow model was developed to determine the heat transfer and hydrodynamic loss of the cooling design. This system model and its predictions for the operating points were validated by pGAPS pre-flight and in-flight measurements~\cite{Okazaki2014}.

\subsection{Time-of-flight\label{s-tof}}
The TOF measures the incoming particle's velocity, provides a high-speed trigger, and serves as a shield/veto for the instrument. A layer of highly segmented plastic scintillator (BC-408 or EJ-200) will completely surround the Si(Li) tracker. Time-of-flight information will be provided by a second layer of scintillator, 1\,m from the top and covering part of the sides. The scintillator will be 0.5\,cm thick and segmented into paddles that are 16\,cm wide and 160--180\,cm long (depending on location). Curved acrylic light guides on both ends of each paddle direct the light to Hamamatsu R7600 photomultiplier tubes (PMTs). Both these PMTs and BC-408 scintillator segments were used successfully on the pGAPS payload. The GAPS scintillators will be thinner than typical particle physics TOF detectors, which are designed to achieve 100--200\,ps time resolution. For GAPS, a time resolution of 500\,ps is sufficient, and so the scintillator can be thin to save weight. Concerning the choice of thickness, it is also important to note that GAPS will trigger on events with $0.2 < \beta < 0.5$, where the light emitted will be 2--6\,times higher than for minimum-ionizing tracks. The timing difference of two PMT signals at opposite ends of the scintillator will provide \texttildelow6\,cm spatial resolution (assuming $v\sub{eff}=15$\,cm/ns in the scintillator). From timing and segmentation, the TOF system will determine the incoming particle trajectory with an angular resolution of \texttildelow5$\,^{\circ}$. 

\section{Conclusions}

The General Antiparticle Spectrometer (GAPS) is a large-acceptance cosmic-ray experiment, designed to measure low-energy antideuterons and antiprotons during Antarctic long-duration balloon (LDB) flights. Cosmic-ray antideuterons are sensitive to a wide range of theoretical models, probing dark matter masses from $\mathcal{O}(10\text{\,GeV})$ to $\mathcal{O}(1\text{\,TeV})$. 

\section{Acknowledgments}
This work is supported in the U.S. by NASA APRA Grants (NNX09AC13G, NNX09AC16G) and the UCLA Division of Physical Sciences and in Japan by  MEXT  grants  KAKENHI  (22340073).   K.  Perez’s work is supported by the National Science Foundation under Award No. 1202958.

%

\providecommand{\href}[2]{#2}\begingroup\raggedright\endgroup

\end{document}